\documentclass[times,12pt]{article}
\usepackage{graphicx}  
\usepackage{authblk} 
\usepackage{bm}        
\usepackage{amssymb}
\usepackage{amsmath} 
\usepackage{epstopdf} 
\usepackage{amsfonts,float}
\usepackage{fancyhdr}
\usepackage{pifont}
\usepackage[hidelinks]{hyperref} 
\usepackage{bbm}
\usepackage{tensor}
\usepackage{mathrsfs}
\usepackage{xcolor}
\usepackage{physics}

\usepackage[titles]{tocloft}
\usepackage{graphicx}

\newcommand{\R}{\mathbb{R}}

\newcommand{\deruno}[2]{\frac{d#1}{d#2}}

\def\beq{\begin{equation}}
\def\eeq{\end{equation}}
\def\ba{\begin{eqnarray}}
\def\ea{\end{eqnarray}}
\def\nn{\nonumber}

\title{REVISITING THE CASIMIR ENERGY WITH GENERAL BOUNDARY CONDITIONS AND APPLICATIONS IN 1D CRYSTALS }

\author[$\dagger$]{ J. M. Mu\~noz-Casta\~neda\footnote{jose.munoz.castaneda@uva.es}}

\author[$\star$]{ M. Bordag}

\author[$\dagger$]{ L. Santamar\'\i a-Sanz}

\affil[$\dagger$]{\footnotesize Departamento de F\'\i sica Te\'orica, At\'omica y \'Optica, Universidad de Valladolid, 47011, Spain}

\affil[$\star$]{\footnotesize Institut f\"ur Theoretische Physik, Universit\"at Leipzig, 04103, Germany}

\date{}

\begin{document}

\maketitle

\begin{abstract}
We obtain new expressions for the Casimir energy between plates that are mimicked by the most general possible boundary conditions allowed by the principles of quantum field theory. This result enables to provide the quantum vacuum energy for scalar fields propagating under the influence of a one-dimensional crystal represented by a periodic potential formed by an infinite array of identical potentials with compact support.

{\it Keywords:}
Quantum field theory under the influence of classical backgrounds; Casimir effect; boundary conditions.

\end{abstract}

\section{Introduction}

As it is well known, two plane parallel homogeneous isotropic $D-1$-dimensional plates in a $D$-dimensional space interacting with a scalar field living between them can be mimicked with boundary conditions over the scalar field \cite{aso-npb13}. The Schr\"odinger eigenvalue problem that characterises the modes of a quantum scalar field living in the space\footnote{We will denote coordinates in $\R\times M $ as $y=(t,x,{\bf y})\in \R\times[0,L]\times\R^{D-1}$. Making this election for the physical space $M$ implies that we assume the plates orthogonal to the $OX$-axis and placed at $x=0$ and $x=L$ respectively.} $M=[0,L]\times\R^{D-1}$ is 
\begin{equation}\label{ec2}
-\Delta_M\psi_\omega(x,{\bf y})=-\left( \Delta_{\R^{D-1}}+\frac{d^2}{dx^2}\right)\psi_\omega(x,{\bf y})=\omega^2\psi_\omega(x,{\bf y}).
\end{equation}
Unitarity of the scalar quantum field theory requires that $-\Delta_M$ has to be selfadjoint and non-negative \cite{aso-npb13}. After separation of variables we arrive to the eigenvalue problem in the orthogonal direction to the plates
\beq\label{ec3}
-\frac{d^2 f_k(x)}{d x^2}=k^2f_k(x),\quad x\in[0,L],
\eeq
meanwhile the eigenvalue problem for the spatial directions parallel to the plates becomes
$
-\Delta_{\R^{D-1}}g_{\bf k}({\bf y})={\bf k}^2g_{\bf k}({\bf y})$, with $ {\bf y},{\bf k}\in\R^{D-1}.
$
Since $\Delta_{\R^{D-1}}$ is selfadjoint, the obstruction for $\Delta_M$ to be selfadjoint is enclosed in $d^2/dx^2$ for $x\in [0,L]$. The most general boundary conditions, including topology changes \cite{aso-npb13}, for the normal modes of the quantum field are in this case given by
\begin{equation}\label{ec5}
\left(
\begin{array}{c}
 f(0)+i\lambda f'(0) \\
 f(L)-i\lambda f'(L) \\
\end{array}
\right)=U\left(
\begin{array}{c}
 f(0)-i \lambda f'(0) \\
 f(L)+i\lambda f'(L) \\
\end{array}
\right)\quad\text{with}\quad U\in {\cal M}_F\subset U(2),
\end{equation}
where $\lambda$ is parameter with units of length that characterises the plates (from now onwards we will assume $\lambda=1$), and ${\cal M}_F$ is the set of $U(2)$ matrices that ensures a non-negative spectrum for the eigenvalue problem \eqref{ec3} (see Ref. \cite{aso-npb13}). Taking for the group $U(2)$ the standard parametrisation
\begin{equation}
U(\alpha,\beta,{\bf n})=e^{i \alpha}\left[\cos(\beta)\mathbb{I}+i\sin(\beta)(\boldsymbol{n}\cdot\boldsymbol{\sigma} )\right],\,\,(\alpha,\beta,{\bf n})\in[0,2\pi]\times[-\frac{\pi}{2},\frac{\pi}{2}]\times S^2,
\end{equation}
being $S^2$ the 2-sphere, and $\boldsymbol{\sigma}=(\sigma_1,\sigma_2,\sigma_3)$ the Pauli matrices. The spectrum of discrete momenta $k$ orthogonal to the plates is given by the zeroes of the function \cite{aso-npb13,jmmc-lmp15}
\ba
h_U(k,L)& =& 2i\operatorname{e}^{i\alpha}\left[((k^2-1)\cos\beta+(k^2+1)\cos\alpha)\sin(kL)-2k\sin\alpha\cos(kL)\right.\nn\\
&-&\left.  2k n_1\sin\beta\right]=-2k(\det(U)-1)\cos(kL)-2 k (U_{12}+U_{21})\nn\\
&+&i\left(  (k^2+1)(\det(U)+1)+(k^2-1){\rm tr}(U)\right)\sin(kL)\label{ec7}.
\ea

\section{Casimir Energy between Plates in arbitrary Dimension}

The general formula for the finite Casimir energy per unit area between two isotropic homogeneous plates mimicked by the boundary conditions \eqref{ec5} is given by,\cite{aso-npb13}
\beq\label{ec8}
\frac{E^{(D)}(U)}{S}=w(D)\frac{L_0^D}{L^D-L_0^D}\int_0^\infty dk\,k^D\left( L-L_0-\frac{d}{dk}\log\left( \frac{h_U(ik,L)}{h_U(ik,L_0)}\right) \right),
\eeq
where $w(D)$ is a global factor just depending on the dimension $D$ of the physical space where the quantum field lives, and $L_0$ a regularisation length that enables to subtract the subdominant divergent terms \cite{aso-npb13}. The Eq.\eqref{ec8} as it appears in Ref. \cite{aso-npb13}  gives the correct answer for the quantum vacuum energy between plates in arbitrary dimension whenever there is no background potential and is valid as well when $L_0>L$. Nevertheless it is necessary to remove the length $L_0$ to provide a formula that can be generalised whenever the operator that characterises the modes in the normal directions has the form
\beq\label{ec9}
{\bf K}=-\frac{d^2}{d x^2}+V(x), \quad x\in[0,L],
\eeq
with $V(x)$ a potential with compact support in $[0,L]$ and no bound states. In order to remove the regularising length $L_0$ we take the limit $L_0\to\infty$ in Eq. \eqref{ec8}. After some straightforward calculations we obtain
\beq\label{ec12}
\lim_{L_0\to\infty}-L_0+\frac{d}{dk}\log\left( h_U(ik,L_0)\right)=\frac{d}{dk}\log\left(\frac{1}{2}(k-i)^2c_U\left(-\frac{k+i}{k-i}\right)\right),
\eeq
where $c_U(z)\equiv \det(U)-{\rm tr(U)}z+z^2=\det(z-U)$. It is of note that 
\beq
\frac{1}{2}(k-i)^2c_U\left(-\frac{k+i}{k-i}\right)=\lim_{L_0\to\infty}\frac{h_U(ik,L_0)}{e^{k L_0}}\equiv h_U^\infty(ik).
\eeq
 Hence, introducing \eqref{ec12} into \eqref{ec8} we obtain a new formula for the Casimir interaction energy per unit area  that is independent of the regularisation length $L_0$:
\beq
\frac{E^{(D)}(U)}{S}=-w(D)\int_0^\infty dk\,k^D\left[ L-\frac{d}{dk}\log\left( h_U(ik,L)\right)
+\frac{d}{dk}\log\left(h_U^\infty(ik)\right)\right]\label{ec13}.
\eeq
This last equation is valid for the case in which there is a potential of compact support between plates that depends only in $x$ as shown in \eqref{ec9}.
\section{Applications to the Quantum Vacuum Energy per Unit Cell for $1D$ Crystals}
Recently the quantum vacuum energy per unit cell for a comb formed by potentials of compact support with no bound states was obtained in Ref. \cite{bor-front19}. In particular it was computed the quantum vacuum energy per unit cell of a scalar quantum field interacting with the background potential
\beq
V(x)=\sum_{n\in\mathbb{Z}}(w_0\delta (x-na)+2w_1\delta'(x-na)),
\eeq
where the $\delta$-$\delta'$potential was defined as in Ref. \cite{gad-pla09}. In Ref. \cite{bor-front19} it is demonstrated that a quantum scalar field propagating along a one-dimensional crystal formed as an array of potentials of compact support smaller than the lattice spacing is equivalent to a scalar quantum field theory in an interval of length given by the lattice spacing, interacting with the potential of compact support that defines the comb centred in the middle point of the interval. In addition, the boundary conditions that the modes of the quantum field must satisfy are given by Bloch semi-periodicity condition, which are quasi-periodic boundary conditions,\cite{bor-front19}
\beq
U_{qp}=\left(
\begin{array}{cc}
 0 & e^{i \theta } \\
 e^{-i \theta } & 0 \\
\end{array}
\right),
\eeq 
where in this situation the parameter $\theta$ is to be interpreted as $\theta=-q L$ with $q$ the quasi-momentum and $L$ the lattice spacing, just after the limit $L_0\to\infty$ is done as in Ref. \cite{bor-front19}. The secular equation that determines the spectrum for the normal modes in this case can be written in terms of the scattering data $\{t(k),r_R(k),r_L(k)\}$ for the potential of compact support when placed over the real line:
\beq
f_\theta(k)\equiv\cos(\theta)-\frac{1}{2t}[e^{-ikL}+e^{ikL}(t^2-r_R r_L)].
\eeq
In this case the limit in \eqref{ec12} replacing $h_U(i k)$ by $f_\theta(i k)$ gives $-\partial_k\log\left( t(ik)\right)$
which enables using straightforward Eq. \eqref{ec13} to obtain a much more physically meaningful formula than the one obtained in Ref. \cite{bor-front19}
\beq
E_0^{fin}(\theta)=-\frac{1}{2\pi}\int_0^{\infty} dk \, k \left[-L +\deruno{}{k} \log \left(f_{\theta}(ik)\right) + \deruno{}{k} \log \left(t(ik)\right)\right].
\eeq
From this last formula, following the previous results from \cite{bor-front19} we can obtain the energy per unit cell of the comb as a summation over the quasi-momentum of the discrete spectra characterised by $f_\theta=0$ :
\beq
E_{\rm comb}^{fin}=\int_{-\pi}^{\pi}\frac{d\theta}{2\pi}E_0^{fin}(\theta).
\eeq
\section{Concluding Remarks}
We have provided a much more general formula for the Casimir energy with general boundary conditions than the one originally obtained in Ref. \cite{aso-npb13}, that is independent of the regularisation length $L_0$. This result enables to extend \eqref{ec13} to much more general situations in which there is a potential of compact support between plates as shown in Eq. \eqref{ec9}. As an application we provided a formula for the quantum vacuum energy per unit cell of a scalar field interacting with a one-dimensional crystal lattice built as an infinite array of identical potentials with compact support. 
\section*{Acknowledgments}

JMMC and LSS acknowledge MINECO (MTM2014-57129-C2-1-P) and Junta de Castilla y Le\'on (BU229P18 and VA137G18).

\end{document}